\documentclass[12pt]{iopart}
\usepackage{iopams}  
\usepackage{graphicx}
\usepackage{cite}
\usepackage[autostyle]{csquotes}
\newcommand{\ket}[1]{|#1\rangle}
\newcommand{\bra}[1]{\langle#1|}
\begin{document}
	
	\title[Interplay of classical and quantum dynamics in a thermal ensemble of atoms]{Interplay of classical and quantum dynamics in a thermal ensemble of atoms}
	
	\author{Arif Warsi Laskar$^1$, Niharika Singh$^2$  \footnote{$^1$ $^,$ $^2$These two authors contributed equally}, Arunabh Mukherjee$^3$, and Saikat Ghosh$^4$}
		
		\address{Department of Physics, Indian Institute of Technology, Kanpur 208016, India}
		\ead{$^4$gsaikat@iitk.ac.in}
		
\begin{abstract}
	
	In a thermal ensemble of atoms driven by coherent fields, how does evolution of quantum superposition compete with classical dynamics of optical pumping and atomic diffusion? Is it optical pumping that first prepares a thermal ensemble, with coherent superposition developing subsequently or is it the other way round: coherently superposed atoms driven to steady state via optical pumping? Using  a stroboscopic probing technique, here we experimentally explore these questions. A 100 $ns$ pulse is used to \textit{probe} an experimentally simulated, \textit{closed} three-level, $\Lambda$-like configuration in rubidium atoms, driven by strong coherent (\textit{control}) and incoherent fields. Temporal evolution of probe transmission shows an initial overshoot with turn-on of \textit{control}, resulting in a scenario akin to lasing without inversion (LWI). The corresponding rise time is dictated by coherent dynamics, with a distinct experimental signature of half-cycle Rabi flop in a thermal ensemble of atoms. Our results indicate that, in fact, optical pumping drives the atoms to a steady state in a significantly longer time-scale that sustains superposed \textit{dark states}. Eventual \textit{control} turn-off leads to a sudden fall in transmission with an ubiquitous signature for identifying \textit{closed} and \textit{open} systems. Numerical simulations and toy-model predictions confirm our claims. These studies reveal new insights into a rich and complex dynamics associated with atoms in thermal ensemble, which are otherwise absent in state-prepared, cold atomic ensembles. 
	
\end{abstract}

				
		\maketitle

\section{Introduction}
  
Engineering superposed quantum state in atomic media has garnered significant interest in past few decades due to its potential as a resource for performing tasks in quantum information \cite{Kimble08,Hammerer10,Reiserer15}. Such engineered states have also led to several dramatic physical effects. For example, electromagnetically induced transparency (EIT) of an otherwise opaque atomic medium \cite{Boller91,Li95,Fleischhauer05} leading to slow and stored light \cite{Hau99,Phillips01,Tanji09} uses a superposed \enquote{dark state}, formed with a strong \textit{control} and a weak \textit{probe} field. With atoms trapped in such states, a small population in the excited state can lead to gain, thereby achieving lasing without (population) inversion (LWI) \cite{Zibrov95,Padmabandu96,Wu08}. Furthermore, being largely immune to any spontaneously scattered photons, superposed states play a pivotal role in generation, storage and retrieval of single-photon Fock states \cite{Duan01,McKeever04,Chou04,Mucke10} or photon pairs \cite{Kuzmich03,Balic05,Kolchin06,Thompson06}, controlled entanglement of ensembles of cold atoms \cite{Lukin00,Matsukevich06,Simon07,Choi08}, teleportation of arbitrary quantum states \cite{Zhang06,Krauter13} and single-photon switches \cite{Turchette95,Shomroni14}. Early experimental demonstration of such superposed states used hot atomic vapors \cite{Bajcsy03,Julsgaard04,Camacho09}, followed soon with usage in several more exotic physical systems, including single-atoms in dipole traps \cite{Darquie05,Streed12} or cold-atomic ensembles and cavities \cite{Reiserer15,Simon07,Ye99,Li13}. While these systems have varied level of experimental complexity, room-temperature atomic vapor remains simplest to implement, with a possible promise of scalability towards a quantum network \cite{Kimble08,Hammerer10}.

In most of these systems, in particular in hot atomic vapors, incoherent (classical) dynamics sets up an intriguing competition with its quantum (coherent) counterpart. For example, in traditional EIT experiments in hot atoms, a transparency window with a sub-natural line-width on a Doppler broadened absorption profile bears a ubiquitous signature of quantum superposition. However, in an \textit{open}, three-level thermal ensemble of atoms, with additional de-cohering channels (figure~\ref{fig:ELD}), one expects several new competing processes to enrich the dynamics  \cite{Fleischhauer05}. Along with time-scales of forming quantum superposition of states (figure~\ref{fig:ELD}(c)), these also include purely classical time-scales of optical pumping of atoms to probe ground state, loss and thermal diffusion  of atoms in and out of light fields (figure~\ref{fig:ELD}(b)). One can therefore ask if there is still useful quantum coherence in the ensemble, how long it takes for such a coherence to build up, and how does it compete with optical pumping and loss. 

Here we explore these questions experimentally in hot rubidium atoms. In particular, we use a stroboscopic technique to probe the transient response of  experimentally simulated \textit{open} and \textit{closed} three-level atoms (figure~\ref{fig:ELD}), driven with coherent (\textit{control}) and incoherent fields. The transmitted peak of the probe pulse at different time instances reveal a rich dynamics, with distinct signatures of classical and quantum regimes. For a \textit{closed} three-level system, we find that with turn-on of the \textit{control} field, there is an initial build up of quantum coherence, resulting in an overshoot in transmission. The corresponding rise-time ($\tau_r$) scales inversely with the control Rabi frequency ($\Omega_c$) i.e. $\tau_r \propto 1/\Omega_c$. Such half-cycle Rabi flops have been observed in laser cooled atoms \cite{Chen98,Echaniz01,Greentree02}. However, to the best of our knowledge, there has not been any prior direct observation of these flops in hot room-temperature atoms. Along with an initial transient coherence which cancels the absorption, a small build up of population in the excited state (also at a rate proportional to $\Omega_c$) leads to an apparent gain peak, akin to LWI \cite{Zibrov95,Padmabandu96,Wu08}. The peak in transparency is followed by an exponential decay to an eventual steady state, with a decay-time ($\tau_d$) that scales as $\tau_d \propto 1/{\Omega_c}^2$. This suggests a steady state, achieved via optical pumping of atoms both to a dark and a non-superposed ground state at a rate $\tau_{OpPump} \sim \frac{1}{\gamma_{eff}}\frac{\Delta_c^2}{\Omega_c^2}$ ($\Delta_c$ representing control field detuning). It is found that $\gamma_{eff} = \gamma_{ex}$, the excited-state life time. We therefore conclude that in hot atomic vapors, at time scales comparable or smaller than the excited state life time, coherence along with a small excited state population leads to LWI. However in the steady state, dark state population builds up over a much longer time scale, set by the detuning and control field strength and governed by optical pumping (i.e for closed vs. open systems). The behavior of the system at control field turn-off reveals the nature of the steady state, along with traces to eventual thermalization of the ensemble. All experimental findings are consistent with numerical modeling

Our observations are consistent with prior work in cold atoms. Transient response has been probed with continuous-wave (c.w.) probe fields \cite{Chen98,Echaniz01,Greentree02} along with observation of coherence build up, to an eventual steady state via optical pumping. However, to the best of our knowledge, such dynamics have remained largely unexplored for hot atomic vapors, for which there is a rich class of dynamical interplay between coherent and incoherent processes. Our stroboscopic technique, with probe pulses at specific time instances, inherently uses a lock-in mechanism, thereby improving the signal-to-noise ratio, as compared to a c.w. probe. Furthermore we find that this probing of the instantaneous response function (as opposed to an integrated response for c.w. field) provides insight into the dynamics of the medium which can be understood using simple toy-models. This study can open up several applications, ranging from pedagogical interests in  testing competing hypothesis (classical or quantum) \cite{Tsang12,Molmer15} to more applied fields of shaping waveforms and controllable photonic switches with hot atomic vapor. 

Next we discuss the experimental methods used in this work. Following this is a detailed comparison of the observed dynamics in various scenarios. We conclude with a discussion on possible implications of these results in hypothesis testing, Zeno like state preparation and pulse shaping.

\section{ Experimental Method:} 

\begin{figure*}
	\centering
	\includegraphics[scale=0.6]{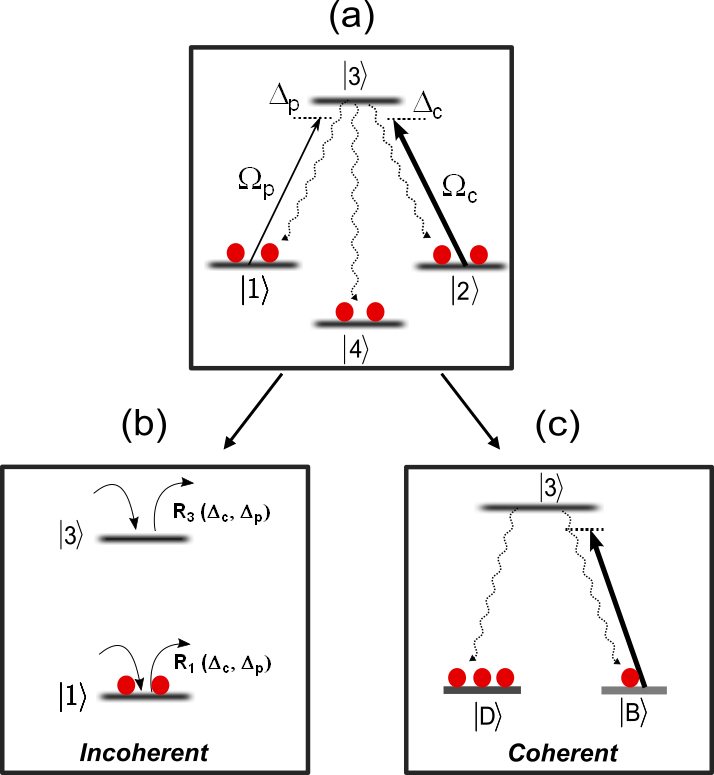}
\caption {\textit{Physical system}: (a) An \textit{open} $\Lambda$ scheme formed with levels  $\ket{1}$, $\ket{2}$ and $\ket{3}$ along with an additional level, $\ket{4}$ accounting for all adjacent states. Transitions $\ket{1}\rightarrow\ket{3}$ and $\ket{2}\rightarrow\ket{3}$ are driven by a weak probe  and a strong control field of Rabi frequencies (detunings): $\Omega_{p}$ ($\Delta_{p}$) and $\Omega_{c}$ ($\Delta_{c}$) respectively. A transparency window in frequency space for such a system will have contributions from: (b) Classical (incoherent) dynamics: If the incoherent pumping rates in and out of the levels $\ket{1}$ and $\ket{3}$, $R_{1}(\Delta_p,\Delta_c)$ and $R_{3}(\Delta_p,\Delta_c)$, are frequency dependent, a transparency window can open up in the midst of absorption profile; and an (c) Quantum dynamics (coherent): Coherent coupling leads to superposed dark ($\ket{D}\equiv(\Omega_{c}\ket{1}-\Omega_{p}\ket{2})/\sqrt{\Omega_{c}^{2}+\Omega_{p}^{2}}$) and bright ($\ket{B}\equiv(\Omega_{c}\ket{2}+\Omega_{p}\ket{1})/\sqrt{\Omega_{c}^{2}+\Omega_{p}^{2}}$) states. Atoms, eventually pumped to dark states (decoupled from excited state) sustains a transparency window. Here we explore how (b) and (c) compete to reach steady state and thermal equilibrium.}
	\label{fig:ELD}
\end{figure*}

For all experiments reported here, we consider $^{85}$Rb $D_{2}$ manifold, with $\ket{1} \equiv \ket{F=3, m_{F}}$,  $\ket{2} \equiv \ket{F=3,m_{F}-2}$ and $\ket{3} \equiv \ket{F^{\prime}=2, m_{F}-1}$ (figure~\ref{fig:ELD}(a)). These levels thereby form a degenerate $\Lambda$ system, with transitions $\ket{1}\rightarrow\ket{3}$ and $\ket{2}\rightarrow\ket{3}$ driven by $\sigma^{-}$ polarized probe field and  $\sigma^{+}$ polarized control field, respectively. Level $\ket{4} \equiv \ket{F=2} $ accounts for all adjacent ground states. Our numerical simulations of a three level $\Lambda$ system reproduce all the observed experimental features well \cite{Li95,Zibrov95,Padmabandu96,Wu08}. A rubidium vapor cell at room temperature, of length $8$ cm and diameter $2$ cm, with Brewster-cut glass windows on either of its side is used as the atomic media. The cell is shielded with three layers of $\mu$-metal sheets along with magnetic coils to cancel any stray magnetic fields.

\begin{figure*}
  		\includegraphics[scale=0.11]{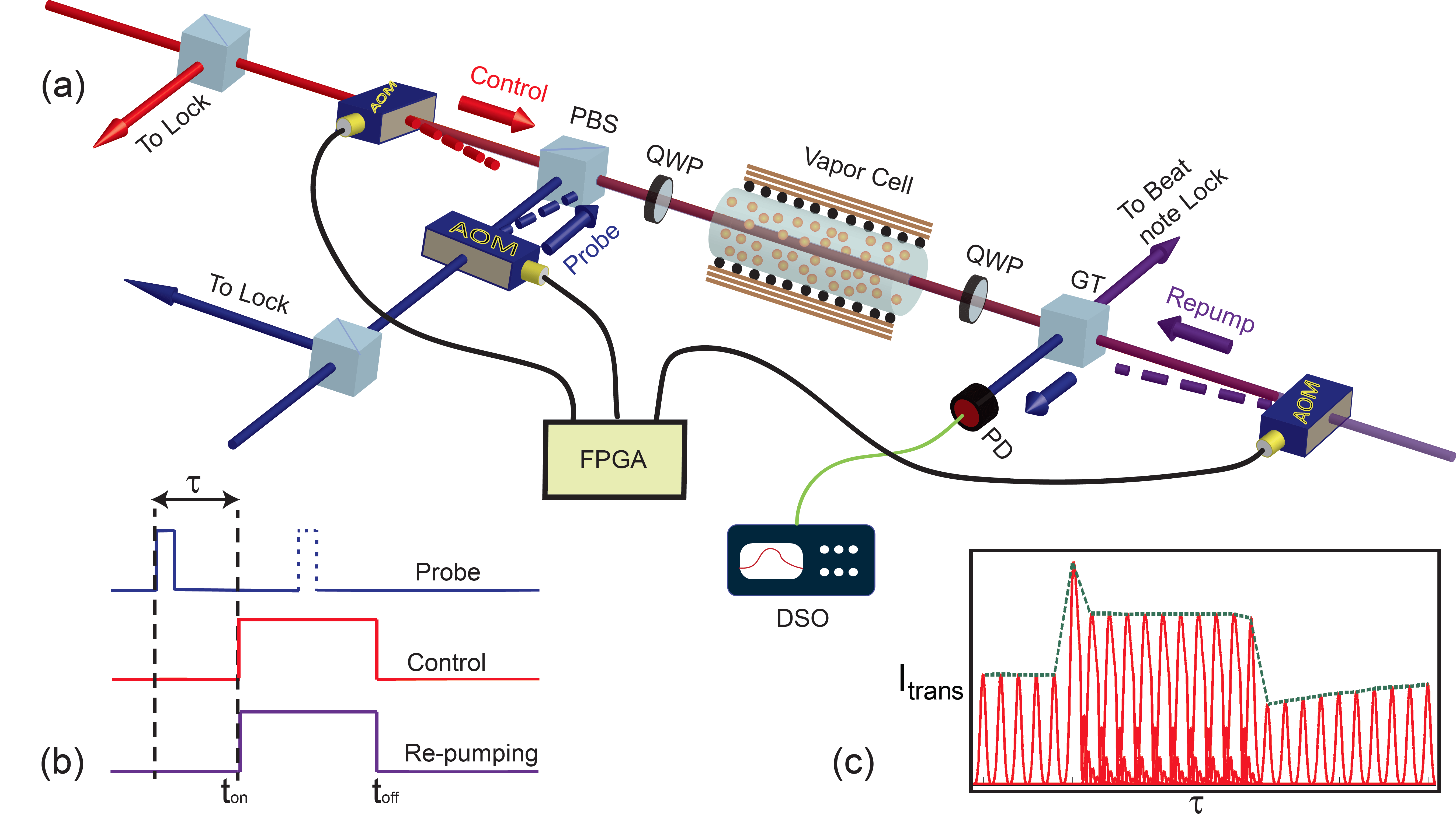} 
  		\caption{\textit{Experimental setup}: (a) A schematic of the experimental setup (details in text). (b) Pulse sequence for the three fields, controlled using a FPGA card. The turn-on and off times of the control and re-pumping fields are $t_{on}= 0$ and $t_{off}=$ 10 $\mu s$ respectively. $\tau$ is the delay between $t_{on}$ and probe turn-on time. (c) A representative plot of stroboscopic probing with the solid red curve depicting the transmitted probe intensity via a series of experiments done with varying $\tau$. The dotted envelope corresponds to a typical experimental transmission profile. Here AOM: acousto-optic modulator, PBS: polarizing beam splitter, QWP: quarter wave-plate, GT: Glan-Thompson polarizing beam splitter, PD: photo detector and DSO: digital storage oscilloscope.}
  		\label{fig:setup}
  	\end{figure*} 

Experimentally, we prepare and compare three physical scenarios: 

(A) A closed three-level system:  To close the system, we use an incoherent re-pumping field. This field pumps the leaked atoms back to the three cycling levels at a rate $1/R$. 

(B) An open three-level system: Corresponds to the usual scenario of EIT experiments in atomic vapor, with just a control and a probe field. 

(C) A system with incoherent pump: A counter-propagating laser is used to saturate the absorption of the probe field, a scenario akin to saturated absorption spectroscopy.

We probe the system in a stroboscopic fashion, recording the time evolution of the transmitted light. The transmitted peak intensity of a pulsed probe (of width 100 ns) is recorded for a series of time instances. The peak power is recorded with a 200 MHz digital oscilloscope, for each time instance $\tau$, the delay between control and probe pulses. A typical transmission profile presented therefore corresponds to an envelope of a series of experiments by varying $\tau$ (figure~\ref{fig:setup}(c)). Experimental timing sequence corresponding to the three fields is shown in figure~\ref{fig:setup}(b).
Control and/or incoherent pump fields are kept on for 10 $\mu s$ (with adiabatic turn on/off $\thicksim$ 200 ns). Repetition time of the entire experiment is 50 $\mu s$ (it takes about 30 $\mu$s for the atomic population to relax back to thermal equilibrium after pump turn-off). 

Digital pulses are programmed with a FPGA (field-programmable-gated-array) card (Opal Kelly XEM3001). Programmed pulses are used to drive acousto-optic modulators (AOM) for turning fields on and off, with a control upto $5$ ns between pulse edges.

Figure~\ref{fig:setup}(a) shows the schematic diagram of the experimental set-up. Control and probe fields are derived from an external cavity diode laser (TOPTICA DL pro) with an output power of $\thicksim$ 90 mW. A second external cavity diode laser is used as the incoherent re-pumping laser. Both lasers operate near $^{85}$Rb $D_{2}$ transition (780 nm) and have a line-width $<<$ 1 MHz. The three beams have a typical diameter of $\thicksim$ 3 mm. A part of both lasers are used for subsequent frequency stabilization and beat-note lock. The control and probe beams are locked 12 MHz blue detuned with respect to $F=3\rightarrow F^{\prime}=2$ transition, while the re-pumping laser is kept near resonance with $F=2\rightarrow F^{\prime}$ transition or $F=3\rightarrow F^{\prime}$ depending on the case under investigation. The fields are pulsed using AOMs at a center rf frequency of 80 MHz. The left and right circularly polarized control and probe fields co-propagate through the Rb vapor cell, while the re-pumping beam which has the same polarization as the probe is sent counter-propagating through it. A Glan-Thompson polarizer is used after the cell for subsequent field separation. Probe beam is then detected on a low noise, amplified photo detector (MenloSystems FPD310-FV) which has a bandwidth of 1.3 GHz. Numerical simulations use the parameters derived from the experiments and the details are described in Appendix.

\section{Observations and discussions}
\bigskip  
\begin{flushleft}
	\textbf{Case A: A \textit{closed} three-level system}
\end{flushleft}

We first consider the case of a \textit{closed} three level system in thermal equilibrium, with equal population in the ground state manifold. Experimentally, we simulate such a \textit{closed} system with an additional counter-propagating re-pumping field (figure~\ref{fig:ELD}(a)), driving the transition $\ket{4}\rightarrow\ket{3}$ and incoherently pumping back atoms, which are otherwise scattered out of the $\Lambda$ system, from $\ket{3}$ to $\ket{4}$. 

\begin{figure*}
 	\includegraphics[scale=0.35]{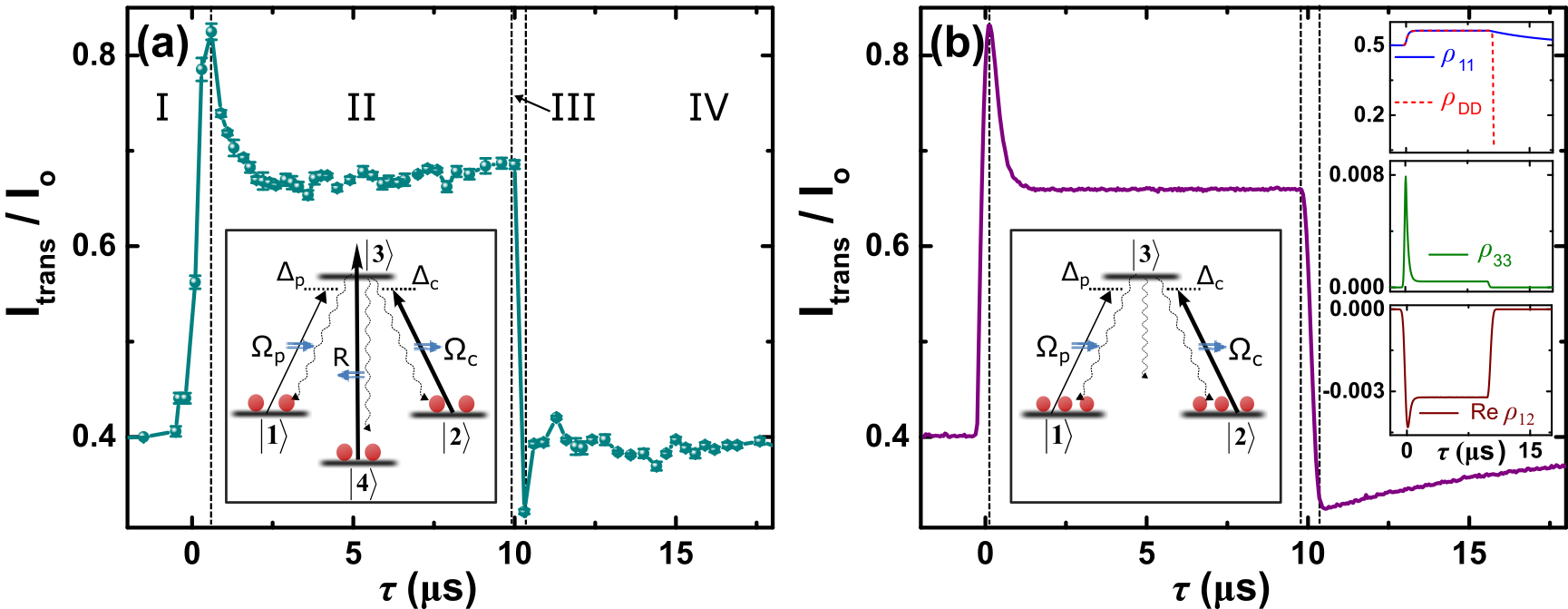} 
 	\caption{\textit{Probe transmission for case(A) (\textit{closed} $\Lambda$ system}): (a) Experimentally recorded peak transmission, with $I_{o}$, $I_{trans}$ being the initial and transmitted probe-peak intensities, respectively. Inset shows energy level configuration of the system. (b) Simulated probe transmission in the absence of re-pumping field. Here $\Omega_{p} = 0.01\gamma_{3}$, $\Omega_{c} = 4.5\gamma_{3}$, $\gamma_{3}=$ 6 MHz, $\gamma_{out}=$ 12 MHz, $\Gamma_{decoh}=$ 0.05 MHz, $\gamma_{th}=$ 0.12 MHz, where $\gamma_{3}=\gamma_{31}+\gamma_{32}=$ 6 MHz and $\rho_{11}^{(0)}=\rho_{22}^{(0)}=\rho_{11}^{eq}=\rho_{22}^{eq}$. These parameters are chosen in accordance with the experiment. Inset shows the energy level configuration, populations in bare states $\ket{1}$ and $\ket{3}$, dark state $\ket{D}$ and  ground state coherence Re$(\rho_{12}$).}
 	\label{fig:closed}
 \end{figure*}

For the numerical modeling (figure~\ref{fig:closed}(b)), we choose a \textbf{\textit{closed}}  three-level system, without any re-pumping field but with a phenomenological decay constant ($\gamma_{out}=$ 12 MHz). The corresponding stroboscopic time-trace is in excellent agreement with the observations. The correspondence between experiment and simulations validates the use of re-pumping field to define a closed three-level system.  

From a typical experimental time trace generated by varying the probe delay $\tau$ (figure~\ref{fig:closed}(a)), one can observe that a steady state is reached following an initial overshoot. Accordingly, the probe transmission can be divided into four distinct time domains: region(I), an initial overshoot region; region (II): decay to a nearly flat steady state region; region(III):control and/or re-pumper turn-off, accompanied with a fast fall in transparency and region(IV): a slow rise back to the initial thermal value (IV). These regions are ubiquitous to all the three cases considered.

		\begin{figure*}
			\includegraphics[scale=0.41]{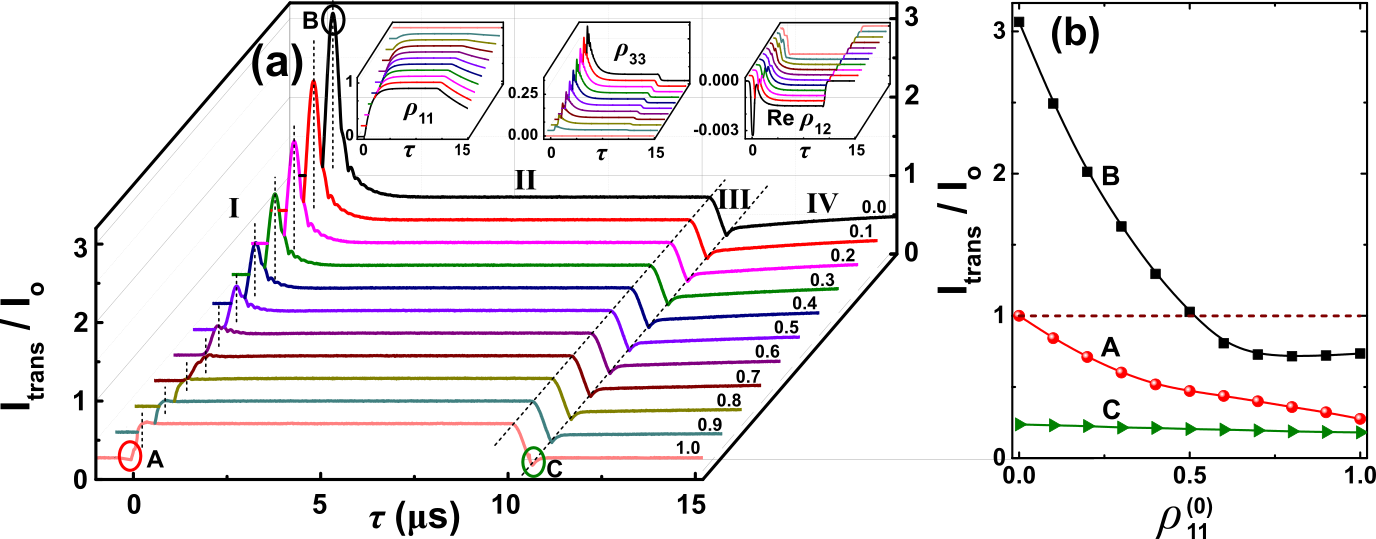}
			\caption{\textit{Evidence of LWI in a closed system}: (a) Simulated probe transmission as a function of initial population in state $\ket{1}$ ($\rho_{11}^{(0)}$) for $\Omega_{c}=5.0\gamma_{3}$. The numbers on top of each curve represent the value of $\rho_{11}^{(0)}$ ($\rho_{22}^{(0)}=1-\rho_{11}^{(0)}$). Remaining parameters are identical to that in figure~\ref{fig:closed}. Inset shows the evolution of populations in bare states $\ket{1}$, $\ket{3}$ and coherence Re$(\rho_{12}$).  (b) Amplitude of probe transmission versus $\rho_{11}^{(0)}$. Here A(red circle): initial $I_{trans}$ without control, B(black square): $I_{trans}$ at overshoot peak in region I (control on) and C(green triangle): $I_{trans}$ at the dip in region III with the control turned off. The region above dotted line depicts the LWI regime.}
			\label{fig:initialcond}
		\end{figure*}

	Figure~\ref{fig:initialcond} shows simulated probe transmission as a function of initial population distribution. In an ideal $\Lambda$ system the dark state is of the form $\ket{D}\equiv(\Omega_{c}\ket{1}-\Omega_{p}\ket{2})/\sqrt{\Omega_{c}^{2}+\Omega_{p}^{2}}$, implying that with all initial population in state $\ket{1}$ ($\rho_{22}^{(0)}$ = 0 and $\Omega_{c}>>\Omega_{p}$), $\ket{D}\thicksim\ket{1}$. Thus at control turn on, the system is already in a dark state and for large fields one achieves steady state transparency almost instantaneously. The rise-time $\tau_r$ is then decided by the control ramp time $\tau_{c}$ when $\Omega_{c}>1/\tau_{c}$ or control Rabi frequency $\Omega_{c}$ when $\Omega_{c}<1/\tau_{c}$. For ideal EIT scenario, one therefore does not expect any overshoot in transmission. 
	
	On the other hand, when $\rho_{22}^{(0)} >\rho_{11}^{(0)}$, one observes a Raman gain, due to transfer of population to state $\ket{3}$, which results in population inversion between $\ket{1}$ and $\ket{3}$. It is intriguing to note that such an overshoot is present only in systems with statistical distribution of initial population in the ground states, and is absent in an otherwise ideal EIT medium with all atoms initially prepared in probe ground state. In simulations, overshoot is not observed for  $\rho_{22}^{(0)}/\rho_{11}^{(0)}<1$.
	
	It can be observed that	the overshoot has a corresponding signature of build-up of Raman coherence in the medium. Numerical simulations confirm a peak in two-photon coherence $\rho_{12}$ (figure~\ref{fig:initialcond}(a)).  Accordingly, one expects this peak to bear signature of two-photon Rabi flop. Furthermore, it can be noted that there is a peak in population of excited state (figure~\ref{fig:initialcond}(a)). The overshoot is therefore akin to LWI,  with a small population in the excited state along with a peak in Raman coherence, $\rho_{12}$ rendering the system transparent.

		 \begin{figure*}
		 	\includegraphics[scale=0.31]{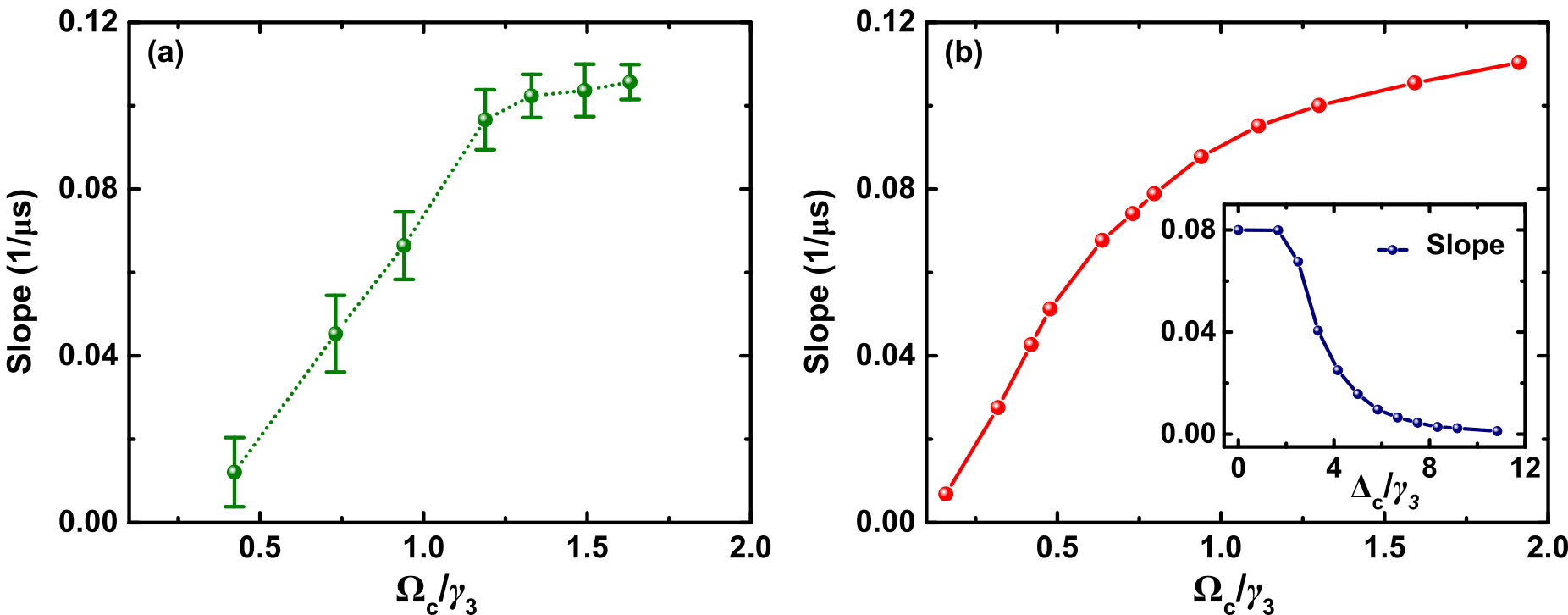}
		 	\caption{\textit{Half cycle Rabi-flop in a \textit{closed} $\Lambda$ system}: (a) Experimental and (b) simulated calculated slope (1/$\tau_r$) of initial rise in region I as a function of $\Omega_{c}$. Inset of frame (b) shows the slope as a function of control detuning $\Delta_{c}$ (=$\Delta_{p}$) for $\Omega_{c} = 5\gamma_{3}$.}
		 	\label{fig:omegac}
		 \end{figure*}

	Interestingly enough, the time scale $\tau_r$  for the rise in region I is measured to be proportional to the rate of two-photon Rabi flop ( with a time-scale $\propto 1/\Omega_{c}$ for a resonant system and $\propto \Delta_{c}/\Omega_{c}\gamma$ for an off-resonant system). Figure~\ref{fig:omegac} shows experimentally observed dependence of rise-time on control Rabi frequency. The initial rise is found to be proportional to 1/$\Omega_{c}$ for small $\Omega_{c}$ and then saturating for $\Omega_{c}>>\Omega_{sat}$ (with $\Omega_{sat}$ being the saturation intensity). The inset of figure~\ref{fig:omegac}(b) numerically confirms that this rise scales as one-photon detuning $\Delta_{c}$ for off-resonant systems (the dependence of probe pulse width and control ramp time on $\tau_r$ is discussed in the Appendix). 
	
	On the contrary if the system is entirely driven by incoherent processes, for e.g. by optical pumping or with an incoherent re-pumping field at a rate R, the rise is found to be much slower. The initial rise time is then $\propto 4\Delta_{c}^{2}/\Omega_{c}^{2}\gamma_{ex}$  and $1/R$ respectively (case C).

	 \begin{figure*}
	 	\centering
	 	\includegraphics[scale=0.35]{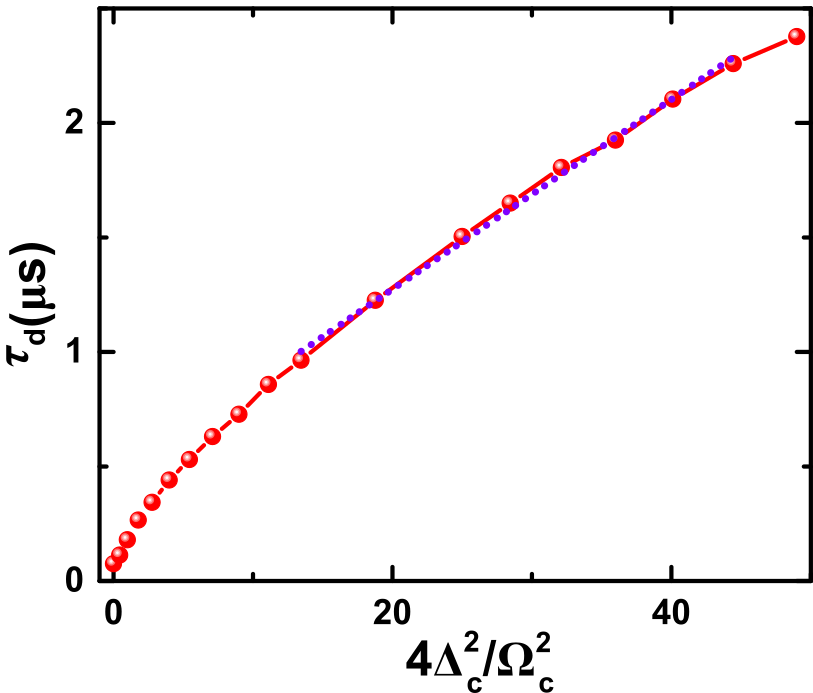}
	 	\caption{\textit{Pumping to steady state in a closed $\Lambda$ system}: Numerically simulated decay time of region II (obtained by exponential fitting) as a function of $\Delta_{c}^{2}/\Omega_{c}^{2}$, with all other parameters identical to that of figure~\ref{fig:closed}. The dotted line shows a linear fit, with slope found close to the excited state decay $\gamma_{ex}=\gamma_{31}+\gamma_{32}+\gamma_{out}$ (24 MHz).}
	 	\label{fig:decay}
	 \end{figure*}  
  	
	In region II, the steady state transparency is achieved via two competing processes: (a) transfer of population to dark state $\ket{D}$ sustaining probe transparency, and (b) incoherent rates populating non superposed state $\ket{1}$ and bright states $\ket{B}$. Accordingly, probe-transparency settles to a lower steady state value than in ideal EIT. The steady state response in general can be expressed as \cite{Ghosh05}:

\begin{equation}
 \rho_{13}^{(ss)} =\frac{-i\Omega_{p}(\rho_{33}^{(ss)}-\rho_{11}^{(ss)})}{2[\gamma_{31}-i\Delta_{p}+\frac{|\Omega_{c}|^{2}/4}{\Gamma_{21}-i(\Delta_{p}-\Delta_{c})}]}\\
+\frac{|\Omega_{c}|^{2}(\rho_{22}^{(ss)}-\rho_{33}^{(ss)})}{4(\gamma_{32}+i\Delta_{c})[\Gamma_{21}-i(\Delta_{p}-\Delta_{c})]}
 \end{equation}

Here superscript (ss) represent steady state value. The first term on r.h.s captures process (a) with coherent state superposition. On the contrary, the second term represents process (b) with incoherent optical pumping in thermal vapor. While region (I) sees a build up of (a), the time scale $\tau_d$  to achieve steady state transparency in region II is set by the optical pumping rate $\thicksim4\gamma_{ex}\Delta_{c}^{2}/\Omega_{c}^{2}$. We verify this dependence in simulations. Figure~\ref{fig:decay} shows a linear dependence of $\tau_d$ on $\Delta_{c}^{2}/\Omega_{c}^{2}$.

At the turn-off of the re-pumping and control fields (regions III and IV), the probe transparency equilibrates back to the initial value (at $\tau_{0}$) in two distinct steps. Initially the transparency falls rapidly to a value much lower than the initial value at $\tau_{0}$. This sudden fall in transparency corresponds to a loss of dark state atoms with the control turn-off, and an adiabatic rotation of the dark-state to a non-superposed state $\ket{1}$ atoms. In simulations, it can be observed that the fall in transmission mimics the decay time of dark state population (inset of figure~\ref{fig:closed}(b)). For a \textit{closed} three level system, since most of the population ends up in the probe ground state $\ket{1}$, this results in a population larger than $\rho_{11}^{(0)}$ and an enhanced probe absorption in region III. The amplitude of this fall thereby gives a direct and alternative measure of number of atoms in the probe ground state. 

Region IV shows a straightforward classical dynamics. The atomic ground state population, in absence of pumping fields, relaxes back to thermal equilibrium. This happens on a time scale set by transit time of atoms ($\thicksim$ 8 $\mu s$) through the probe beam diameter. 

One can thereby conclude that the transient evolution in the closed $\Lambda$ system is dominated by coherent evolution in regions I and III while incoherent processes dictate the dynamics in regions II and IV.

\begin{figure*}
	\includegraphics[scale=0.33]{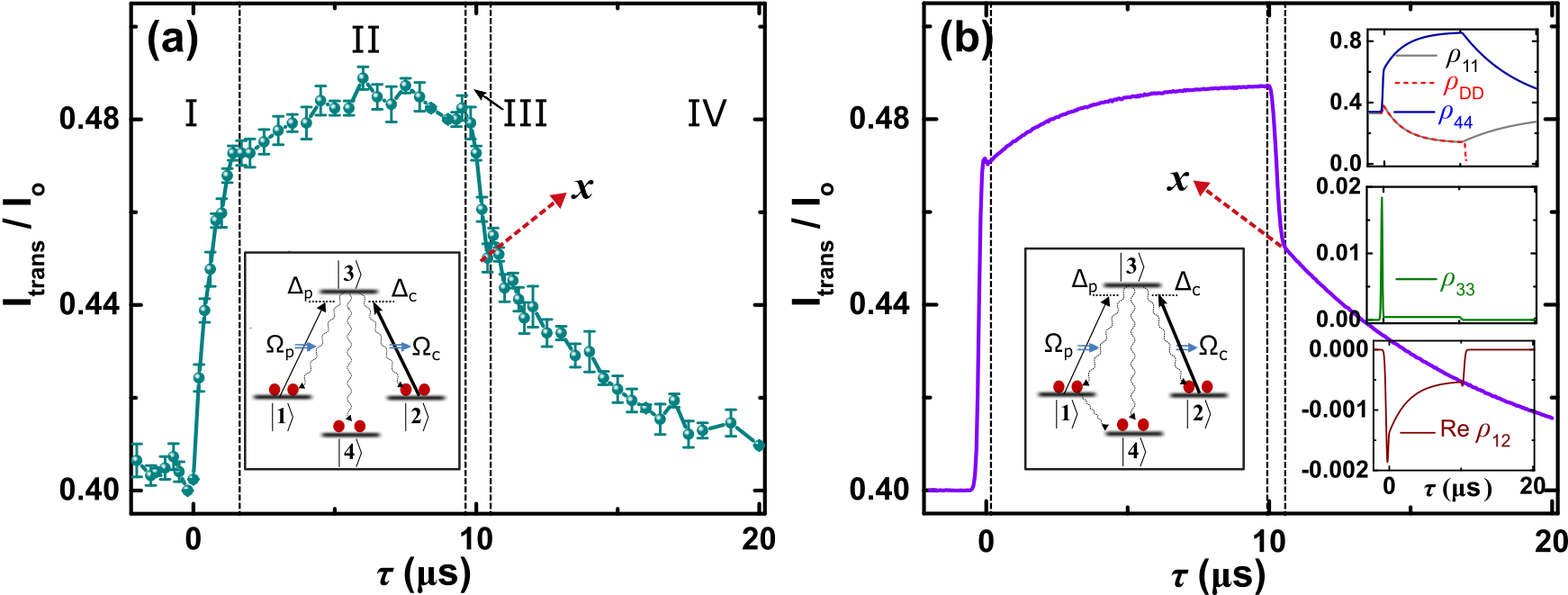}
	\caption{\textit{Probe transmission for case(B) (an open $\Lambda$ system)}: (a) and (b) depict experimental and simulated time traces, respectively. For simulations, an additional probe ground state decay term of rate $ \sim 0.01 \Omega_{c}$ is used with $\Omega_{c} = 7 \gamma_{3}$. These parameters are chosen in accordance with the experiment. Inset shows energy level configuration, populations in states $\ket{1}$, $\ket{3}$ and $\ket{4}$, dark state $\ket{D}$ and  ground state coherence Re$(\rho_{12})$. When the control is turned off, the sharp and partial fall of probe transmission (to point $X$) is due to adiabatic rotation of dark state atoms to bare states. From $X$ the population relaxes back to initial thermal equilibrium within few transit times.}
	\label{fig:open}
\end{figure*}

 \bigskip  
 \begin{flushleft}
 	\textbf{Case B: An open three-level system}
 \end{flushleft}
 
 To compare the observations of a \textit{closed} system with an \textit{open} three-level system, we next consider a scenario without the re-pumping field. Figure~\ref{fig:open} shows typical experimental probe transmission  for such a system (figure~\ref{fig:open}(a)), together with the corresponding numerical simulation (figure~\ref{fig:open}(b)). Due to the presence of adjacent excited levels, there is an off-resonant scattering out of $\ket{1}$. In numerical simulations, we account for this by explicitly adding a small incoherent pump-out of  ground state $\ket{1}$, at a rate $\thicksim$ 0.01 $\Omega_{c}$ (inset of figure~\ref{fig:open}(b)), which is less than $\Omega_c^2\gamma_{ex}/\Delta_c^2\thicksim$ 0.05 $\Omega_c$ for $\Delta_c\thicksim$ 50 MHz ($F^{\prime}=3$).
 
  Once again, one can divide the time-trace into four distinct regions. Akin to Case (A), build up of coherence build is accompanied with a sharp rise in region I, though an overshoot is washed out. The rise time again scales as $1/\Omega_c$. Following the fast rise, the system reaches a steady state through optical pumping in dark, bright states as well as atoms out of the ground-state manifold.

When the control is turned off, probe transparency equilibrates in a starkly different fashion as compared to that in \textit{closed} system. As discussed, the initial sharp fall (until point \textit{x} in figure~\ref{fig:open}) is due to atoms trapped in the dark state and adiabatically rotated to bare state $\ket{1}$. However unlike case (A), the corresponding drop is not all the way down. At the instance of control turn-off a significant fraction of the population is also in state $\ket{4}$, which is not a part of the closed system. The magnitude of the fall region III can therefore be quantified as an indicator of effective population in $\ket{1}$, and therefore of dark state. 

The system revives to thermal equilibrium value within the transit time-scale in region IV. It can also be noted that the fall amplitude in region III can be controlled by changing the incoherent pump rate.

  \begin{figure*}
  	\includegraphics[scale=0.35]{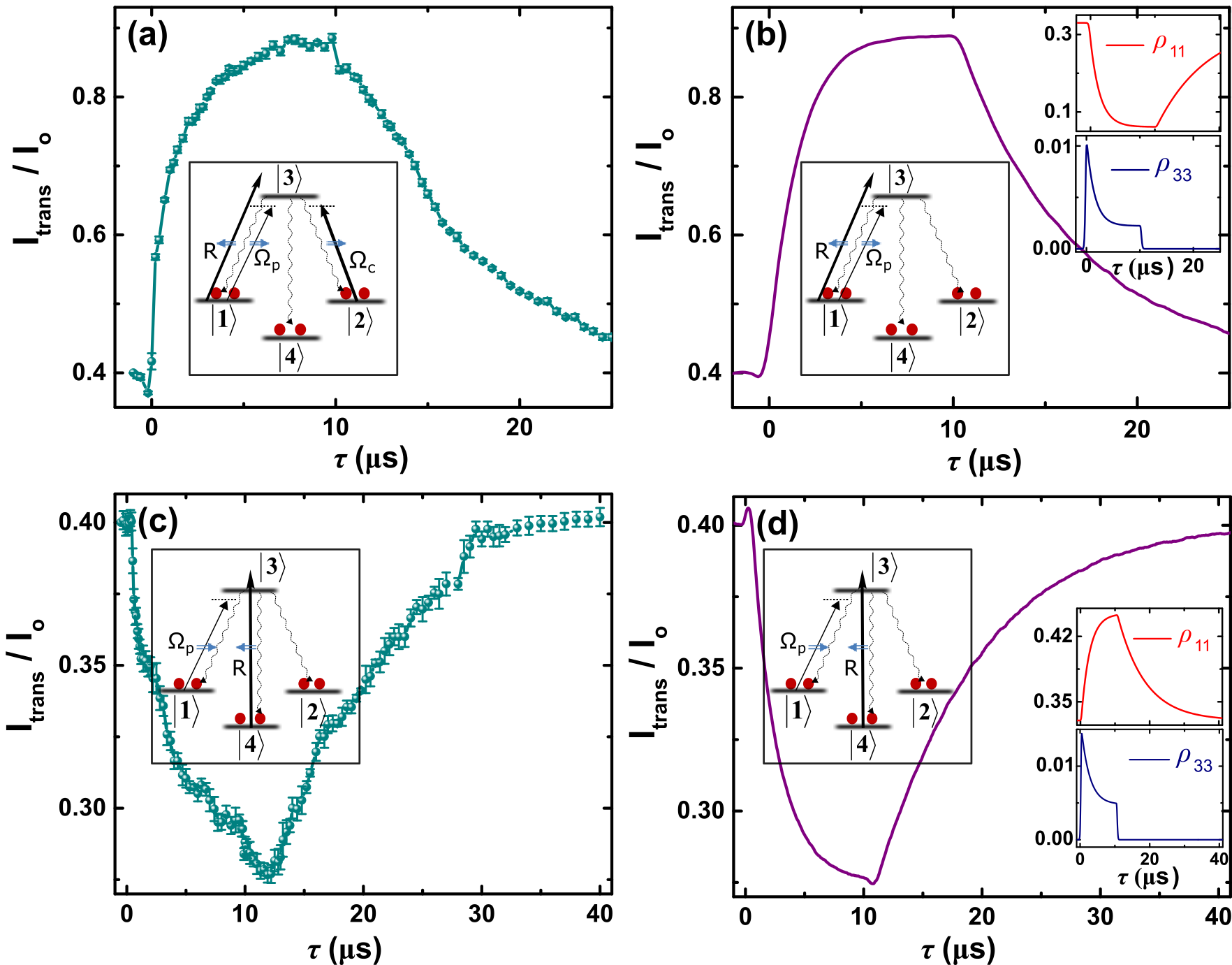}
  	\caption{\textit{Probe transmission for case(C) (incoherent dynamics)}:  Frames (a) and (b) show the experimental and numerical plots of probe transmission when counter-propagating re-pumping field is driving the same transition as the probe. Here R $\thicksim \Omega_{c}\thicksim$ 0.3 $\gamma_{3}$. Simulations are in absence of control field as shown in frame (b). Frames (c) and (d) show experimental and simulations without control but with re-pumping field, driving transition $\ket{4}\rightarrow\ket{3}$. Here R $\thicksim$ 0.3 $\gamma_{3}$. Rest of the parameters are same as in figure~\ref{fig:closed}. The numerical simulation parameters are chosen in accordance with the experiment. Inset show the respective energy level configuration.}
  	\label{fig:incoherent}
    \end{figure*} 
  
  \bigskip  
  \begin{flushleft}
  	\textbf{Case C: An incoherently pumped three-level system}
  \end{flushleft}
  
 Here we consider two simple test cases dominated by incoherent dynamics. Figure~\ref{fig:incoherent}(a) and (b) correspond to the first case, with an incoherent counter-propagating re-pumping field on resonance with the same transition as the probe ($\ket{1}\rightarrow\ket{3}$), along with the control field driving $\ket{2}\rightarrow\ket{3}$. Near resonance, both probe and re-pumping fields address same velocity group of atoms with the strong re-pumping field saturating the transition and depleting the atoms from the probe ground state. Along with the control induced EIT, this incoherent pumping further increases the probe beam transparency. The rise is comprised of two competing time scales: time-scale for build up of coherence and that for saturation effects due to re-pumping. We observe incoherent effects to dominate, with a slow rise time ($\thicksim$ 1 $\mu s$) set by the population transfer rates. Numerical simulations, purely with incoherent pump and in complete absence of control field capture most of the essential features (figure~\ref{fig:incoherent}(b)) with a rise time ($\tau_r$) of probe transmission is $\propto 1/R$. At the turn-off of the control and re-pumping fields, the probe transparency returns to its initial value within a transit-time scale. 

 An exact opposite probe response is obtained in the absence of control field and an incoherent re-pumping beam, driving the transition $\ket{4}\rightarrow\ket{3}$ (figures~\ref{fig:incoherent}(c)). With re-pump field turned-on, most of the population is optically pumped to the probe ground state thereby increasing the probe absorption. At turn-off, the probe absorption revives to its initial value within the transit-time scale. The slow fall ($\thicksim$ 4 $\mu s$) and rise times ($\thicksim$ 9 $\mu s$) are completely dictated by the incoherent part of the dynamics. The corresponding numerical simulation (figure~\ref{fig:incoherent}(d)) is in good agreement with the observation. 
 
 It can also be noted that at re-pump turn-off, any sudden fall of transparency is not observed. This is in accordance with a complete absence of dark-states in the system.

 \section{Conclusion}
 
Evolution of an open quantum system (with a state $\hat{\rho}$) can in general be described with a master equation $\dot{\hat{\rho}}(t)= -i[\hat{H},\hat{\rho}] +\hat{L}(\hat{\rho})$, where the Hamiltonian $\hat{H}$ is responsible for coherent dynamics while the Lindblad operator $\hat{L}$ corresponds mostly to incoherent processes. In this paper we present a study on the interplay of such coherent and incoherent phenomena in controlling the response of a thermal population of three-level atomic system driven by a strong control and a weak probe. An incoherent counter-propagating re-pumping laser provides an additional handle to control the dynamics of the system by either making the system closed or by saturating the probe transition. From this perspective, Case(A) then corresponds to a scenario dominated by $\hat{H}$, Case(C) dominated by $\hat{L}$ while Case(B) captures an intermediate regime. Interestingly enough, for the intermediate case (B), we find that a sharp and partial fall of probe-transparency at control field turn-off bears a unique signature of the relative interplay between $\hat{H}$ and $\hat{L}$. While for Case(A), the fall in transparency was all the way, for Case(C) there was hardly any. For competing classical and quantum theories, such a ubiquitous signature can provide a universal benchmark. There has been considerable recent interest in testing competing hypothesis, describing the same physical system \cite{Tsang12,Molmer15}. The measurements described here can thereby provide a platform for hypothesis testing in future. 

Furthermore, the overshoot of transparency at early times due to fast build up of coherence provides an intriguing possibility of freezing the ensemble to a state of large quantum coherence using repeated, projective measurements. Such Zeno-like state preparation can significantly improve coherent properties of hot-atomic ensemble. It can also be noted that these results are not necessarily limited to atomic systems. For example, in complex solid-state materials \cite{Lodahl15} or confined atoms in hollow core fiber \cite{Ghosh05}, with a myriad of incoherent processes competing with quantum superpositions, our technique can aid in identifying effective quantum superpositions generated in such systems. Finally, the controllable time traces hold promise of several applications, including pulse amplification with LWI, shaping waveforms and designing optical switches with controlled slew-rates.

\bigskip

\textbf{\textit{Acknowledgement:}} SG acknowledges support from IIT-K(initiation grant), DST-SERB(SB/S2/LOP-05/2013). Authors acknowledge Amar Bhagwat, Bimalendu Deb, Harshawardhan Wanare, Luat T. Vuong and Mishkatul Bhattacharya for stimulating discussions and careful reading of the manuscript.

\appendix
\setcounter{equation}{0}
\renewcommand{\theequation}{A.\arabic{equation}}
\renewcommand\thefigure{A.\arabic{figure}}    
\setcounter{figure}{0}    

\section*{Appendix}

\begin{flushleft}
\textit{\textbf{Numerical Simulations}}
\end{flushleft}

For the numerical experiments, we use coupled Maxwell-Bloch equations to time evolve the probe pulse in the presence of changing control field.
The Bloch part of the dynamics is governed by a master equation (in the interaction representation)  $\dot{\rho(t)}= -i[\hat{H},\rho] +\hat{L}(\rho)$. Here the first term on the right accounts for coherent interactions while the second term  $\hat{L}$ represents  irreversible incoherent processes in the system. Such processes include relaxations in and out of the closed three-level system, transit time effects and population transfer due to an additional incoherent re-pumping field. The inclusion of transit time in the density matrix accounts for the effect of thermal velocity distribution of atoms to first order. We do not perform any additional Doppler averaging of the resulting simulated curves.

The effective Hamiltonian (under usual dipole and rotating wave approximation) can be expressed as

 \begin{eqnarray}
 \hat{H}& =(\Delta_{c}-\Delta_{p})\ket{2}\bra{2}-\Delta_{p}\ket{3}\bra{3} \nonumber \\
 &-\frac{\Omega_{p}(z,t)}{2}\ket{1}\bra{3}-\frac{\Omega^{\ast}_{p}(z,t)}{2}\ket{3}\bra{1}
 -\frac{\Omega_{c}(z,t)}{2}\ket{2}\bra{3}
 -\frac{\Omega^{\ast}_{c}(z,t)}{2}\ket{3}\bra{2}
 \end{eqnarray}

 \bigskip
 
Here the control Rabi frequency is defined as $\Omega_{c}(z,t)\thicksim\Omega_{c}(t)=\Omega_{c}(0)e^{-(t-t_{on})^2/2\tau_c^{2}}$ for $t\leq t_{on}$, $\Omega_{c}(0)$ for $t_{on}<t<t_{off}$ and $\Omega_{c}(0)e^{-(t-t_{off})^2/2\tau_c^{2}}$ for $t\geq t_{off}$, where we have assumed that the control field undergoes negligible absorption, remaining unchanged along the propagation length. Here $\tau_c$ is the control ramp time. The turn-on and turn-off times of the control field are $t_{on}= 0$ and $t_{off}=$ 10 $\mu s$ respectively. 

The pulsed probe, of width $\tau_p$, has a corresponding Rabi frequency, $\Omega_{p}(z,t)=\Omega_{p} (z,0)e^{-(t-\tau)^2/2\tau_p^{2}}$. The probe turn-on time, $\tau$, is varied from shot to shot . The initial Rabi frequencies are defined as  $\Omega_{c}(0)=d_{c}\centerdot\varepsilon_{c}(z)/\hbar$, $\Omega_{p}(z,0)=d_{p}\centerdot\varepsilon_{p}(z)/\hbar$, with $d_{c}$ and $d_{p}$ being the transition dipole moments between states $\ket{2}$ and  $\ket{3}$, and $\ket{1}$ and $\ket{3}$ respectively. $\varepsilon_{c}(z)$ and $\varepsilon_{p}(z)$ are the electric field amplitudes of the corresponding fields. The detunings of these lasers from the corresponding atomic transitions are $\Delta_{c}=\omega_{32}-\omega_{c}$ and $\Delta_{p}=\omega_{31}-\omega_{p}$, where $\omega_{p}$ and $\omega_{c}$ correspond to carrier frequency of probe and control pulses. 

The evolution of atomic states according to the time dependent master equation is given by the following set of equations:

\begin{eqnarray}
\frac{\partial \rho_{11}}{\partial t}=&i \frac{\Omega_{p}}{2}(\rho_{31}-\rho_{13})+\gamma_{31}\rho_{33}
-\Gamma_{th}(\rho_{11}-\rho^{eq}_{11})\\
\frac{\partial \rho_{22}}{\partial t}=&i \frac{\Omega_{c}}{2}(\rho_{32}-\rho_{23})+\gamma_{32}\rho_{33}-\Gamma_{th}(\rho_{22}-\rho^{eq}_{22}) \\
\frac{\partial \rho_{33}}{\partial t}=&i\frac{\Omega_{p}}{2}(\rho_{13}-\rho_{31})+i \frac{\Omega_{c}}{2}(\rho_{23}-\rho_{32})+\gamma_{ex}\rho_{33}-\Gamma_{th}(\rho_{33}-\rho^{eq}_{33})\\
\frac{\partial \rho_{44}}{\partial t}=&\gamma_{out}\rho_{33}-\Gamma_{th}(\rho_{44}-\rho^{eq}_{44})\\
\frac{\partial \rho_{12}}{\partial t}=&i \frac{\Omega_{p}}{2}\rho_{32}-i \frac{\Omega_{c}}{2}\rho_{13}-[\Gamma_{12}+\Gamma_{th}+i(\Delta_{c}-\Delta_{p})]\rho_{12}\\
\frac{\partial \rho_{13}}{\partial t}=&i \frac{\Omega_{p}}{2}(\rho_{33}-\rho_{11})-i \frac{\Omega_{c}}{2}\rho_{12}-[\Gamma_{13}+\Gamma_{th}-i\Delta_{p}]\rho_{13}\\
\frac{\partial \rho_{23}}{\partial t}=&i \frac{\Omega_{c}}{2}(\rho_{33}-\rho_{22})-i \frac{\Omega_{p}}{2}\rho_{21}-[\Gamma_{23}+\Gamma_{th}-i\Delta_{c}]\rho_{24}
\end{eqnarray}

where
\begin{eqnarray}
\Gamma_{12}&=&\Gamma_{decoh}\\
\Gamma_{13}&=&\Gamma_{decoh}+\frac{\gamma_{ex}}{2}\\
\Gamma_{23}&=&\Gamma_{decoh}+\frac{\gamma_{ex}}{2}\\
\gamma_{ex}&=&\gamma_{31}+\gamma_{32}+\gamma_{out}
\end{eqnarray}

\begin{figure*}
	\includegraphics[scale=0.35]{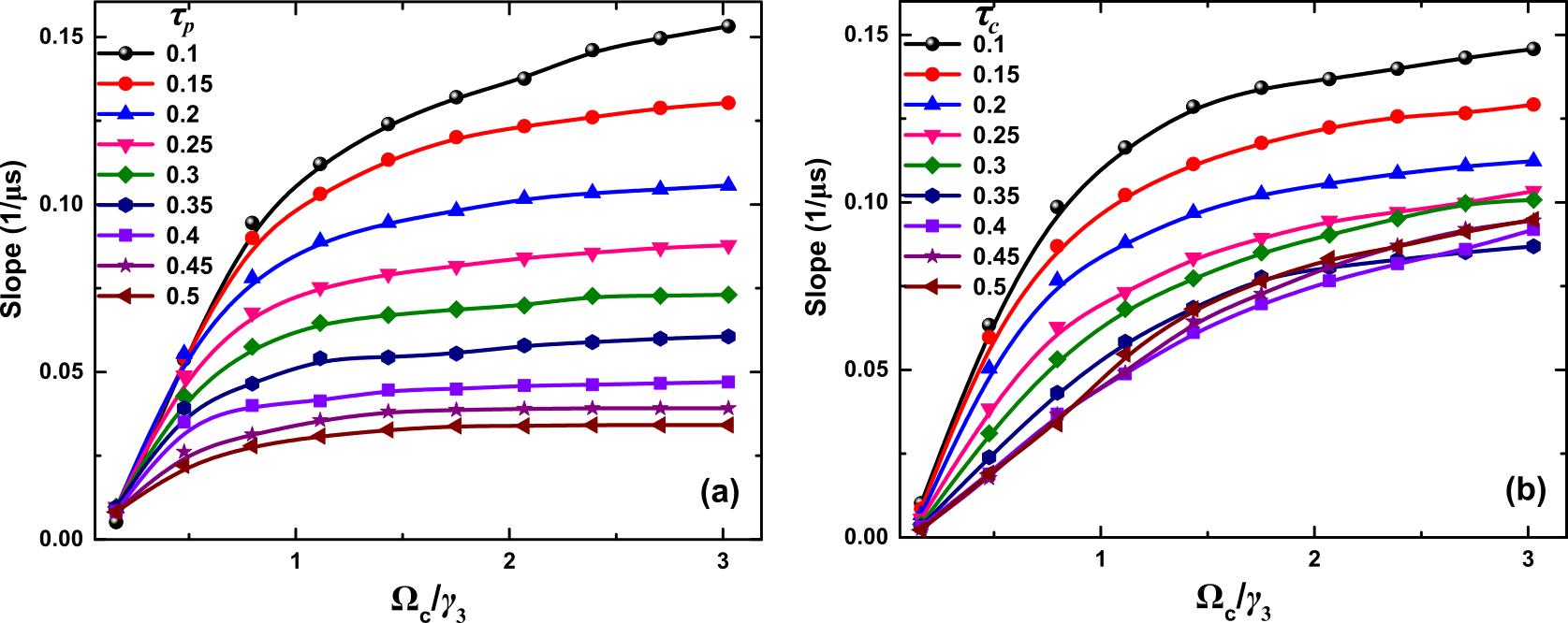}
	\caption{\textit{Effect of probe pulse width ($\tau_{p}$) and control ramp time ($\tau_{c}$) in a closed $\Lambda$ system}: Frames (a) and (b) show the numerically obtained slope (1/$\tau_r$) of initial rise region I as a function of control Rabi frequencies for varying $\tau_{p}$ and $\tau_{c}$ respectively. The numbers in front of the legends correspond to the value of $\tau_{p}$ and $\tau_{c}$ in $\mu s$. Remaining parameters are same as in figure~\ref{fig:closed}.}
	\label{fig:ramp}
\end{figure*}	

The incoherent pump terms are incorporated as population transfer rates in the density matrix equations, without any change in the coherence terms.  Here $\gamma_{31(32)}$ is the radiative decay from from the excited state $\ket{3}$ to $\ket{1}$ ($\ket{2}$). $\gamma_{out}, \Gamma_{decoh}$ and $\Gamma_{th}$ are the radiative decay from $\ket{3}$ out of the closed $\Lambda$ system, decoherence rate and transit time decay respectively.

The above equations are numerically integrated to determine the induced polarization, $\rho_{13}$, as seen by the probe. 
Maxwell wave propagation equation in turn determines the fields

\begin{equation}
\frac{1}{c}\frac{\partial \Omega_{p}}{\partial t}+\frac{\partial \Omega_{p}}{\partial z}=-i \mu \rho_{13}(z,t)
\end{equation} which is integrated self-consistently. Here $\mu=Nd_{p}^{2}\omega_{p}/\hbar\epsilon_{0}$ where $c$ and $\epsilon_{0}$ correspond to speed of light and dielectric susceptibility in vacuum  respectively.  The transmitted probe pulse is $\Omega_{p}(t+\tau)=\Omega_{p}(t)+\int_{0}^{L}\alpha Im(\rho_{13}(z,t))dz$, where $\textit{L}$ is the propagation length in the cell and $\alpha$ is a constant.

\bigskip

\begin{flushleft}
	\textit{\textbf{Effect of pulse width and rise time}}
\end{flushleft}

	Figure~\ref{fig:ramp} shows the dependence of control ramp time $\tau_{c}$ and probe pulse width $\tau_{p}$ on the slope of rise region I in a closed system (figure~\ref{fig:closed}). It is observed that for a fixed control strength, the slope (1/$\tau_r$) is inversely proportional to $\tau_p$. For slow turn on of control pulse, the rise slope is limited by $\tau_{c}$ and does not show any change with control strength at large $\Omega_{c}$.  For $\tau_{c}>1/\gamma_3$, the saturation happens at much larger $\Omega_{c}$.

  \bigskip 
  \bigskip

   	\begin{flushleft}
   		\textbf{References}
   	\end{flushleft}

\end{document}